\documentstyle[prl,aps,epsbox]{revtex}
\begin{document}
\draft
\twocolumn[\hsize\textwidth\columnwidth\hsize\csname@twocolumnfalse%
\endcsname 
\title{Chaos Induced by Quantization}
\author{
Takaomi Shigehara, Hiroshi Mizoguchi, Taketoshi Mishima
}
\address{
Department of Information and Computer Sciences,  
Saitama University,    
Shimo-Okubo, Urawa, Saitama 338-0825, Japan
}
\author{
Taksu Cheon
}
\address{
Laboratory of Physics, Kochi University of Technology,
Tosa Yamada, Kochi 782-8502, Japan\\
}
\date{January 28, 1998}
\maketitle
\begin{abstract}
In this paper, we show that two-dimensional billiards 
with point interactions inside exhibit a chaotic nature 
in the microscopic world, although their classical 
counterpart is non-chaotic. 
After deriving the transition matrix of the system by 
using the self-adjoint extension theory 
of functional analysis, 
we deduce the general condition for the appearance of 
chaos. 
The prediction is confirmed by numerically examining 
the statistical properties of energy spectrum 
of rectangular billiards with multiple point interactions inside. 
The dependence of the level statistics 
on the strength as well as the number 
of the scatterers is displayed. 

\vspace*{3mm}
\end{abstract}
%
%
]
\narrowtext

\section{Introduction} 

The billiard is a suitable tool for 
examining generic features of dynamical systems. 
This is because, in spite of its simplicity, 
it covers the wide range of dynamical behaviors,  
going from the most regular (integrable) to 
the most irregular (chaotic) depending on its shape.  
Pseudointegrable \cite{RB81} billiards discussed in this paper 
is obtained by putting point scatterers into an 
integrable billiard. 
Their dynamical behavior 
is somewhat trivial in classical mechanics.  
A classical point particle never 
hits a point scatterer. More precisely, 
the set of trajectories of a point particle which 
hit point scatterers is of measure zero in the 
classical phase space. 
Thus we may neglect point obstacles in classical 
mechanics. 
However, the situation is drastically changed 
in quantum mechanics. Owing to the 
uncertain principle, the point particle gains 
a ``size'' and as a result, there might be 
a possibility of its hitting point scatterers. 
Our purpose is to show this is in fact the case 
and to clarify the condition for its occurrence. 
In order to emphasize that 
such phenomena originate from quantum effects, 
they might be called {\em wave chaos} \cite{AS91}. 

A particular case of a single scatterer 
located at the center of a rectangular billiard 
has been discussed in details elsewhere 
\cite{SH94}--\cite{SM97}. 
We generalize the formulation and derive 
the transition matrix in case of multiple scatterers.   
Based on this, a new numerical evidence for 
the existence of wave chaos is exhibited. 
We observe that the ``degree'' of chaos 
strongly depends on the number of the scatterers. 

The paper is organized as follows. 
In Sect.2, we briefly mention the measure of chaos 
in quantum mechanics, which is described by 
statistics of energy levels. 
The quantum-mechanical 
treatment of point scatterers is discussed in 
Sect.3. In spite of seeming simplicity of the problem, 
a somewhat complicated argument based on self-adjoint 
extension is needed. 
In Sect.4, we clarify the condition for the appearance 
of wave chaos and its validity is confirmed by 
numerical experiments with a rectangular billiard 
in Sect.5. 
The paper is summarized in Sect.6. 

\section{Quantum-Mechanical Measure of Chaos}

Although no mathematical proof exists, 
it is widely conjectured \cite{BO89} 
that the statistical properties 
of quantum energy spectrum directly reflect the dynamical 
nature of the corresponding classical system. 
Generic integrable systems obey Poisson statistics, 
while generic chaotic systems are described by 
the predictions of the Gaussian orthogonal ensembles 
(GOE) \cite{ME67}. 
One of such statistical measures is the nearest-neighbor 
level spacing distribution $P(S)$, which is defined 
such that $P(S)dS$ gives the probability to find 
the spacing between any two neighboring energy eigenvalues 
in the interval $(S,S+dS)$. 
Depending on the dynamical property of classical system, 
$P(S)$ takes a form 
\begin{eqnarray}
\label{eq2-1}
P(S) = \left\{
\begin{array}{ll}
\displaystyle
\exp (-S), & \mbox{(integrable)}, \\[0.5ex]
\displaystyle
\frac{\pi S}{2} \exp (-\frac{\pi S^2}{4}), & \mbox{(chaotic)}. 
\end{array} \right.
\end{eqnarray}
The degeneracy of eigenvalues frequently happens in integrable 
systems, while it hardly occurs in chaotic systems. 
Another convenient statistics is the spectral rigidity 
\begin{eqnarray}
\label{eq2-2}
\hspace*{-5ex}
\Delta_3 (L) =  \left\langle
\min_{A,B} \frac{1}{L} \int_{E-L/2}^{E+L/2} 
\hspace{-4ex}
\left( N(E')-AE'-B \right)^2 dE'
\right\rangle_{E},
\end{eqnarray}
where $N(E)$ is the staircase function which counts 
the number of eigenstates below energy $E$. 
The brackets $\langle \cdot \rangle_E$ signify  
the average with respect to the energy 
in the energy region under consideration. 
The $\Delta_3 (L)$ statistics represents the average of 
the least square deviation of the staircase function $N(E)$ 
from the best straight line fitting it 
in some intervals of length $L$. 
It takes a form 
\begin{eqnarray}
\label{eq2-3}
\Delta_3 (L) = 
\left\{
\begin{array}{ll}
\displaystyle
\frac{L}{15}, & \mbox{(integrable)}, \\[1ex]
\displaystyle
\frac{1}{\pi^2}(\ln L - 0.0687), 
& \mbox{(chaotic)},  
\end{array} \right.
\end{eqnarray}
depending on the dynamical property. 

\section{Formulation}

We first consider a quantum point particle of mass $M$ 
moving freely in a two-dimensional bounded region (billiard) $S$. 
Let us denote the area of $S$ by the same symbol. 
We impose the Dirichlet boundary condition so that wave functions 
vanish on the boundary of $S$. 
The eigenvalues and the corresponding normalized eigenfunctions 
are denoted by $E_n$ and $\varphi _n (\vec {x})$ respectively; 
\begin{eqnarray}
\label{eq3-1}
H_0 \varphi _n(\vec{x}) \equiv -{\nabla^2 \over {2M}} 
\varphi _n(\vec{x})= E _n \varphi _n(\vec{x}). 
\end{eqnarray}
The Hamiltonian $H_0$ is the kinetic operator 
in $L^2(S)$ with domain $D(H_0)=H^2(S)\cap H^1_0(S)$ 
in terms of the Sobolev spaces. 
The Green's function of $H_0$ is written as 
\begin{eqnarray}
\label{eq3-2}
G^{(0)}(\vec{x},\vec{y};\omega ) = 
\sum\limits_{n=1}^\infty 
{{{\varphi _n(\vec{x})\varphi _n(\vec{y})} 
\over {\omega -E_n}}}, 
\end{eqnarray}
where $\omega$ is the energy variable. 
The average level density of the system is given by 
\begin{eqnarray}
\label{eq3-3}
\rho_{av} = \frac{MS}{2\pi}, 
\end{eqnarray}
which is energy-independent. 

We now place a single point scatterer at 
$\vec{x}_1$ inside the billiard. 
The most naive manner for this purpose is to 
define the scatterer in terms of the Dirac's delta function;  
\begin{eqnarray}
\label{eq3-4}
H=H_0+ v \delta^{(2)} (\vec{x}-\vec{x}_1).
\end{eqnarray}
However, the Hamiltonian $H$ is not mathematically sound. 
This can be seen from the eigenvalue equation of $H$, 
which is reduced to 
\begin{eqnarray}
\label{eq3-5}
\sum_{n=1}^{\infty}
\frac{\varphi_n(\vec{x}_1)^2}{\omega-E_n}=v^{-1}.  
\end{eqnarray}
Since the level density is proportional to 
$E^{(d-2)/2}$ in spatial dimension $d$, 
the infinite series in Eq.(\ref{eq3-5}) 
does not converge for $d \geq 2$. 

To handle the divergence, a scheme for renormalization is called for.  
The most general scheme is given by the self-adjoint extension 
theory of functional analysis \cite{AG88}. 
We first consider in $L^2(S)$ the nonnegative operator 
\begin{eqnarray}
\label{eq3-6}
H_{\vec{x}_1} = - \left. {\nabla^2 \over {2M}} 
\right|_{C^{\infty}_0 (S-\vec{x}_1)}  
\end{eqnarray}
with its closure ${\bar H_{\vec{x}_1}}$ in $L^2(S)$.   
Namely, we restrict $D(H_0)$ to the functions 
which vanish at the location of the point scatterer; 
\begin{eqnarray}
\label{eq3-7}
D(\bar{H}_{\vec{x}_1}) =  
\{ \psi (\vec{x}) \in D(H_0) \vert 
\psi (\vec{x}_1)=0  \}.  
\end{eqnarray}
By using integration by parts, 
it is easy to prove that 
the operator ${\bar H_{\vec{x}_1}}$ is symmetric (Hermitian). 
But it is not self-adjoint. Indeed, the equation 
for the adjoint of ${\bar H_{\vec{x}_1}}$ 
\begin{eqnarray}
\label{eq3-8}
{\bar H_{\vec{x}_1}}^* \psi_{\omega}(\vec{x}) = 
\omega \psi_{\omega}(\vec{x}), 
\ \ \ \psi_{\omega} \in D({\bar H_{\vec{x}_1}}^*),   
\end{eqnarray}
has a solution for $Im \ \omega \neq 0$ \cite{ZO80}; 
\begin{eqnarray}
\label{eq3-9}
\psi_{\omega}(\vec{x}) =  
G^{(0)}(\vec{x},\vec{x}_1;\omega ), 
\ \ \vec{x} \in S-\vec{x}_1, 
\end{eqnarray}
indicating 
\begin{eqnarray}
\label{eq3-10}
\hspace*{-5ex}
D(\bar{H}_{\vec{x}_1}^*) & 
= & 
D(\bar{H}_{\vec{x}_1}) \oplus N(\bar{H}_{\vec{x}_1}^* -\omega) 
\oplus N(\bar{H}_{\vec{x}_1}^* -\bar{\omega}) \nonumber \\
& \neq & 
D(\bar{H}_{\vec{x}_1}),  
\end{eqnarray}
where $N(A)$ is the kernel of an operator $A$. 
Equation (\ref{eq3-10}) means that 
${\bar H_{\vec{x}_1}}$ has the deficiency indices $(1,1)$ 
and as a result,  
${\bar H_{\vec{x}_1}}$ has one-parameter family of 
self-adjoint extensions $H_{\theta_1}$; 
\begin{eqnarray}
\label{eq3-11}
D(H_{\theta_1}) & = & 
\{ f \vert 
f = \varphi + c  ( \psi_{i\Lambda} 
- e^{i\theta_1} \psi_{-i\Lambda} ); \nonumber \\[1ex]
&& \varphi\in D(\bar{H}_{\vec{x}_1}), c\in {\bf C}, 
0 \leq \theta_1 < 2\pi \}, \nonumber \\[1ex]
H_{\theta_1} f & = &
\bar{H}_{\vec{x}_1} \varphi+i \Lambda c 
( \psi_{i\Lambda} + e^{i\theta_1} \psi_{-i\Lambda} ),   
\end{eqnarray}
where $\Lambda >0$ is a scale mass. 
With the aid of Krein's formula \cite{AG88}, 
we can write down 
the Green's function for the Hamiltonian $H_{\theta_1}$ as 
\begin{eqnarray}
\label{eq3-12}
\begin{array}{l}
\displaystyle
G_{\theta_1}(\vec{x},\vec{y};\omega) = 
G^{(0)}(\vec{x},\vec{y};\omega) 
\\[1ex]
\displaystyle
+ \mbox{} G^{(0)}(\vec{x},\vec{x}_{1};\omega)
T_{\theta_1}(\omega)
G^{(0)}(\vec{x}_1,\vec{y};\omega). 
\end{array}
\end{eqnarray}
In Eq.(\ref{eq3-12}), 
the transition matrix $T_{\theta_1}$ is calculated by 
\begin{eqnarray}
\label{eq3-13}
\hspace*{-5ex}
T_{\theta_1}(\omega) =\frac{1-e^{i\theta_1}}
{(\omega-i\Lambda) c_{i\Lambda}(\omega)-
e^{i\theta_1}(\omega+i\Lambda) c_{-i\Lambda}(\omega)}, 
\end{eqnarray}
where
\begin{eqnarray}
\label{eq3-14}
\hspace*{-5ex}
c_{\pm i\Lambda}(\omega)=
\int_{S}
G^{(0)}(\vec{x},\vec{x}_1;\omega)
G^{(0)}(\vec{x},\vec{x}_1;\pm i\Lambda)d\vec{x}. 
\end{eqnarray}
Using the resolvent equation, we have 
\begin{eqnarray}
\label{eq3-15}
T_{\theta_1}(\omega)=(v_{1}^{-1}-G(\omega))^{-1}, 
\end{eqnarray}
where 
\begin{eqnarray}
\label{eq3-16}
v_{1}^{-1}=\Lambda \cot \frac{\theta_1}{2}
\sum_{n=1}^{\infty}
\frac{\varphi_{n}(\vec{x}_1)^{2}}{E_{n}^2+\Lambda^{2}}, 
\end{eqnarray}
\vspace*{-6mm}
\begin{eqnarray}
\label{eq3-17}
G(\omega)=\sum_{n=1}^{\infty}
\varphi_{n}(\vec{x}_{1})^{2}
(\frac{1}{\omega-E_{n}}+\frac{E_{n}} {E_{n}^2+\Lambda^{2}}).
\end{eqnarray}
The constant $v_{1}$ is a coupling strength of the point scatterer, 
the value of which ranges over the whole real number   
as one varies $0\leq\theta_1 <2\pi$.  
It follows from Eq.(\ref{eq3-15}) that 
the eigenvalues of $H_{\theta_1}$ are determined by 
\begin{eqnarray}
\label{eq3-18}
G(\omega)=v_{1}^{-1}.
\end{eqnarray}
On any interval $(E_m,E_{m+1})$, 
the function $G$ is monotonically decreasing, ranging 
over the whole real number. 
This means that the eigenvalue equation (\ref{eq3-18}) has 
a single solution $\omega_m$ on each interval for any $v_{1}$. 
The eigenfunction of $H_{\theta_1}$ corresponding to 
an eigenvalue $\omega_m$ is given by 
\begin{eqnarray}
\label{eq3-19}
\hspace*{-3ex}
\psi_m (\vec{x}) \propto G^{(0)}(\vec{x},\vec{x}_1;\omega_m) 
= \sum_{n=1}^{\infty}
\frac{\varphi_n (\vec{x}_1)}{\omega_m - E_n} \varphi_n (\vec{x}). 
\end{eqnarray}

We proceed to the case of multiple ($N \geq 1$) point scatterers. 
Let us denote the position of the $k$-th scatterer by $\vec{x}_k$. 
In this case, we should start with the operator 
\begin{eqnarray}
\label{eq3-20}
H_{X} = - \left. {\nabla^2 \over {2M}} 
\right|_{C^{\infty}_0 (S- X)}  
\end{eqnarray}
and its closure ${\bar H_{X}}$ in $L^2(S)$. 
Here we set $X=\{\vec{x}_1,\cdots,\vec{x}_N\}$. 
The operator ${\bar H_{X}}$ is symmetric 
(but not self-adjoint) and the equation 
\begin{eqnarray}
\label{eq3-21}
\bar{H}^*_{X} \psi_{\omega}(\vec{x}) = 
\omega \psi_{\omega}(\vec{x}), \ \ \ \psi_{\omega} \in 
D(\bar{H}^*_{X}), 
\end{eqnarray}
has the $N$ independent solutions for $Im \ \omega \neq 0$;   
\begin{eqnarray}
\label{eq3-22}
\psi^{(k)}_{\omega}(\vec{x}) =  
G^{(0)}(\vec{x},\vec{x}_k;\omega ), \ \ \vec{x} 
\in S-X, 
\end{eqnarray}
$k=1,\cdots,N$. 
This indicates that 
${\bar H_{X}}$ has the deficiency indices 
$(N,N)$ and as a result, it has $N^2$-parameter family of 
self-adjoint extensions in general;  
\begin{eqnarray}
\label{eq3-23}
\hspace*{-5ex}
D(H_{U,X}) & = & 
\{ f \vert f = \varphi + \sum_{k=1}^{N} c_k ( \psi^{(k)}_{i\Lambda} 
+ \sum_{l=1}^{N} U_{kl} \psi^{(l)}_{-i\Lambda} ) ; \nonumber \\ 
\hspace*{-5ex}
&& \varphi\in D(\bar{H}_{X}), c_k \in {\bf C} \}, \nonumber \\  
\hspace*{-5ex}
H_{U,X} f & = &
\bar{H}_{X} \varphi+i \Lambda \sum_{k=1}^{N} c_k ( \psi^{(k)}_{i\Lambda} 
- \sum_{l=1}^{N} U_{kl} \psi^{(l)}_{-i\Lambda} ), 
\end{eqnarray}
where $U_{ij}$ denotes an arbitrary $N$-dimensional unitary matrix. 
Using Krein's formula, we can relate the Green's function of $H_{U,X}$ 
to $G^{(0)}$; 
\begin{eqnarray}
\label{eq3-24}
\begin{array}{l}
\displaystyle
G_{U,X}(\vec{x},\vec{y};\omega) = 
G^{(0)}(\vec{x},\vec{y};\omega) \\
\displaystyle
+ \mbox{} \sum_{k,l=1}^{N} G^{(0)}(\vec{x},\vec{x}_{k};\omega)
T(\omega)_{kl}
G^{(0)}(\vec{x}_{l},\vec{y};\omega),  
\end{array} 
\end{eqnarray}
where the transition matrix $T(\omega)$ should satisfy 
\begin{eqnarray}
\label{eq3-25}
\begin{array}{l}
\displaystyle 
[T(\omega)]^{-1}_{kl} - [T(\omega')]^{-1}_{kl} \\[1ex]
\displaystyle 
= G^{(0)}(\vec{x}_k,\vec{x}_l;\omega') -
  G^{(0)}(\vec{x}_k,\vec{x}_l;\omega).   
\end{array}
\end{eqnarray}
Equation (\ref{eq3-25}) indicates that it is sufficient to define 
the transition matrix for some fixed $\omega$ since 
then $T(\omega)$ for any $\omega$ follows from Eq.(\ref{eq3-25}). 
Also note that Eq.(\ref{eq3-24}) implies 
\begin{eqnarray}
\label{eq3-26}
T(\omega)^{\dagger} = T(\bar{\omega}). 
\end{eqnarray}
The general form of $T(\omega)$ which satisfies Eqs.(\ref{eq3-25}) and 
(\ref{eq3-26}) is given by 
\begin{eqnarray}
\label{eq3-27}
T(\omega)^{-1} = A-G(\omega),  
\end{eqnarray}
where 
\begin{eqnarray}
\label{eq3-28}
\hspace{-5ex}
G(\omega)_{kl} = \left\{
\begin{array}{l}
\displaystyle 
\sum_{n=1}^{\infty}
\varphi_{n}(\vec{x}_{k})^{2}
(\frac{1}{\omega-E_{n}}+\frac{E_{n}} {E_{n}^2+\Lambda^{2}}), \\ 
\hspace{25ex} (k=l), \\
\displaystyle 
G^{(0)}(\vec{x}_k,\vec{x}_l,\omega), \hspace{3ex} (k\neq l), 
\end{array} \right.
\end{eqnarray}
and $A$ is any $N$-dimensional Hermitian matrix. 
If $A$ is not diagonal, different scatterers $\vec{x}_k$ 
are connected by the boundary conditions. On the other hand, 
if $A$ is diagonal the point scatterers are independent, namely, 
one has separated boundary conditions at each $\vec{x}_k$. 
We here restrict ourselves to the latter which is important 
from a practical point of view. In this case, 
$A$ can be parameterized as 
\begin{eqnarray}
\label{eq3-29}
A_{kl} = \delta_{kl} v_k^{-1}, \hspace{3ex} v_k \in {\bf R}.  
\end{eqnarray}
One can regard $v_k$ as the strength of the $k$-th scatterer. 
The eigenvalue of $H_{U,X}$ is determined by the poles of $T(\omega)$; 
\begin{eqnarray}
\label{eq3-30}
det ( T(\omega)^{-1} ) = 0, 
\end{eqnarray}
which is reduced to Eq.(\ref{eq3-18}) for $N=1$. 

A final remark is that one can show the relation 
\begin{eqnarray}
\label{eq3-31}
U = -^{t}[T(-i\Lambda)T(i\Lambda)^{-1}].  
\end{eqnarray}
Equation (\ref{eq3-31}) indicates that the unitarity of $U$ 
is equivalent to the fact that $T(i\Lambda)$ is a normal matrix. 

\section{Condition for Strong Coupling}

In this section, we discuss the condition for strong coupling under which 
point scatterers have a substantial effect on energy spectrum 
of the empty billiard. 
For this purpose, we examine the energy-dependence of the inflection 
points of $G$ in Eq.(\ref{eq3-17}), based on a semi-quantitative 
argument suitable for examining the statistical properties of spectrum. 

We first consider the case of a single scatterer. 
The first notice is that the average value of   
$\varphi_n (\vec{x}_1)^2$ among many $n$ 
is independent of the energy; 
\begin{eqnarray}
\label{eq4-1}
\left\langle \varphi_n (\vec{x}_1)^2 \right\rangle_n \simeq 1/S. 
\end{eqnarray}
We thus recognize from Eq.(\ref{eq3-19}) that 
if $\omega_m \simeq E_m$ (resp. $E_{m+1}$) for some $m$, 
then $\psi_m \simeq \varphi_m$ (resp. $\psi_m \simeq \varphi_{m+1}$).   
This implies that a point scatterer distorts the wave function 
if the eigenvalue $\omega_m$ is located around the midpoint of 
the interval $(E_m,E_{m+1})$. 
For such $\omega_m$, 
the value of $G$ can be estimated by the principal integral 
as follows,  
since the contributions on the summation of $G$ from 
the terms with $n \simeq m$ cancel each other;   
\begin{eqnarray}
\label{eq4-2}
G(\omega_m) \simeq g(\omega_m), \ \ \ 
\omega_m \simeq \frac{E_{m}+E_{m+1}}{2}, 
\end{eqnarray} 
\begin{eqnarray}
\label{eq4-3}
\hspace{-5ex}
g(\omega) & = &  
\left\langle \varphi_{n}(\vec{x}_{1})^{2} \right\rangle_n \rho_{av}
P  \int_{0}^{\infty} \left( 
\frac{1}{\omega-E}+\frac{E}{E^2+\Lambda^2}
\right) dE \nonumber \\
\hspace{-5ex}
& = & \frac{M}{2\pi}\ln \frac{\omega}{\Lambda}. 
\end{eqnarray} 
Here we have defined a continuous function $g(\omega)$ 
which behaves like 
an interpolation of the inflection points of $G$. 
Equation (\ref{eq4-2}) indicates that the wave function 
mixing mainly occurs in the eigenstate with an eigenvalue 
such that 
\begin{eqnarray}
\label{eq4-4}
g(\omega_m) \simeq v_1^{-1}. 
\end{eqnarray} 

The ``width'' of the strong coupling region 
(allowable error of $v_1^{-1}$ in Eq.(\ref{eq4-4}))
is estimated by considering a linearized eigenvalue equation. 
Expanding $G$ at 
$\omega_m \simeq (E_{m}+E_{m+1})/2$, 
we can rewrite the eigenvalue equation as 
\begin{eqnarray}
\label{eq4-5}
G'(\omega_m)(\omega-\omega_m) \simeq v^{-1}-G(\omega_m). 
\end{eqnarray} 
In order that Eq.(\ref{eq4-5}) has a solution 
$\omega\simeq \omega_m$, the range of RHS has to be restricted to 
\begin{eqnarray}
\label{eq4-6}
\left| v^{-1} - G(\omega_m) \right| \ \ \alt \ \
\frac{\Delta}{2},  
\end{eqnarray} 
\vspace*{-1ex}
\begin{eqnarray}
\label{eq4-7}
\Delta \equiv   
\left| G'(\omega_m) \right| 
\rho_{av}^{-1}, 
\end{eqnarray} 
where we have defined the width $\Delta$ which 
is nothing but the variance of the linearized $G$ 
on the interval $(E_m,E_{m+1})$. 
The magnitude of $\left| G'(\omega_m) \right|$ 
can be estimated as follows;  
\begin{eqnarray}
\label{eq4-8} 
\left| G'(\omega_m) \right| 
& = & \sum_{n=1}^{\infty} \left(
\frac{\varphi_{n}(\vec{x}_{1})}
{\omega_m-E_{n}} \right)^2 \nonumber \\
& \simeq & \left\langle \varphi_{n}(\vec{x}_{1})^{2} \right\rangle_n  
\sum_{n=1}^{\infty} 
\frac{2}
{\{(n-\frac{1}{2})\rho_{av}^{-1} \}^2} \nonumber \\
& = &   
8 \left\langle \varphi_{n}(\vec{x}_{1})^{2} \right\rangle_n  
\rho_{av}^2 
\sum_{n=1}^{\infty} \frac{1}{(2n-1)^2} \nonumber \\
& = & 
\pi^{2} \left\langle \varphi_{n}(\vec{x}_{1})^{2} \right\rangle_n  
\rho_{av}^2. 
\end{eqnarray}
The second equality follows from the approximation that 
the unperturbed eigenvalues are distributed with a mean 
interval $\rho_{av}^{-1}$ in the whole energy region. 
Noticing Eqs.(\ref{eq3-3}) and (\ref{eq4-1}), 
we obtain 
\begin{eqnarray}
\label{eq4-9}
\Delta \simeq \frac{\pi M}{2},  
\end{eqnarray} 
which is independent of the energy $\omega$. 
We can summarize the findings as follows; 
The effect of a point scatterer of coupling strength $v_1$ 
is substantial mainly in the eigenstates with eigenvalue $\omega$ 
such that 
\begin{eqnarray}
\label{eq4-10}
\left| v_1^{-1} - \frac{M}{2\pi} \ln \frac{\omega}{\Lambda}\right| \alt 
\frac{\Delta}{2} \simeq  \frac{\pi M}{4}. 
\end{eqnarray}

The condition (\ref{eq4-10}) is generalized to the case of 
multiple scatterers. 
Equation (\ref{eq4-1}), which is the main assumption 
in the above argument, is valid even for the perturbed 
eigenfunction (\ref{eq3-19}). 
This allows us to repeat the above 
when the number of the point scatterers increases. 
We conclude that the $k$-th point scatterer with 
strength $v_k$ affects the energy spectrum in the 
energy region such that 
\begin{eqnarray}
\label{eq4-11}
\left| v_k^{-1} - \frac{M}{2\pi} \ln \frac{\omega}{\Lambda}\right| \alt 
\frac{\pi M}{4}, \hspace{3ex} k=1,\cdots,N. 
\end{eqnarray}

It is worthy to emphasize that the strong coupling region 
shifts with a logarithmic dependence of energy. 
This indicates that 
the system recovers the integrability in the semiclassical 
(high-energy) limit for any strength $v_k$. 

\section{Numerical Example}

\begin{table}[b]
\caption{Location of the point scatterers}  
\begin{center}
{\normalsize
\begin{tabular}{|r|c|c|} \hline
\multicolumn{1}{|c|}{$k$} &
\multicolumn{1}{c|}{$x_{k}$} &
\multicolumn{1}{c|}{$y_{k}$} \\ \hline
  1  &  0.6224826 &  0.2758356  \\ \hline 
  2  &  0.8202505 &  0.4561782  \\ \hline 
  3  &  0.1802603 &  0.6365209  \\ \hline 
  4  &  0.3780281 &  0.8168635  \\ \hline 
  5  &  0.5757960 &  0.2332624  \\ \hline 
  6  &  0.7735638 &  0.4136051  \\ \hline 
  7  &  0.1335736 &  0.5939477  \\ \hline 
  8  &  0.3313415 &  0.7742904  \\ \hline 
  9  &  0.5291093 &  0.1906893  \\ \hline 
 10  &  0.7268772 &  0.3710320  \\ \hline 
\end{tabular}
}
\end{center}
\label{table1}
\end{table}

In this section, we set the scale mass $\Lambda=1$ 
without losing generality. 
This makes all physical quantities dimensionless. 
We consider a quantum particle with mass $M=2\pi$ moving in 
a rectangular billiard with side-lengths 
$(l_{x},l_{y})=(\pi/3, 3/\pi)$, and hence $S=1$. 
The average level density is $\rho_{av}=1$, 
according to Eq.(\ref{eq3-3}). 
The eigenvalue $E_{n_x,n_y}$ and the corresponding eigenfunction 
$\varphi_{n_x,n_y}$ are given by 
\begin{figure}[t]
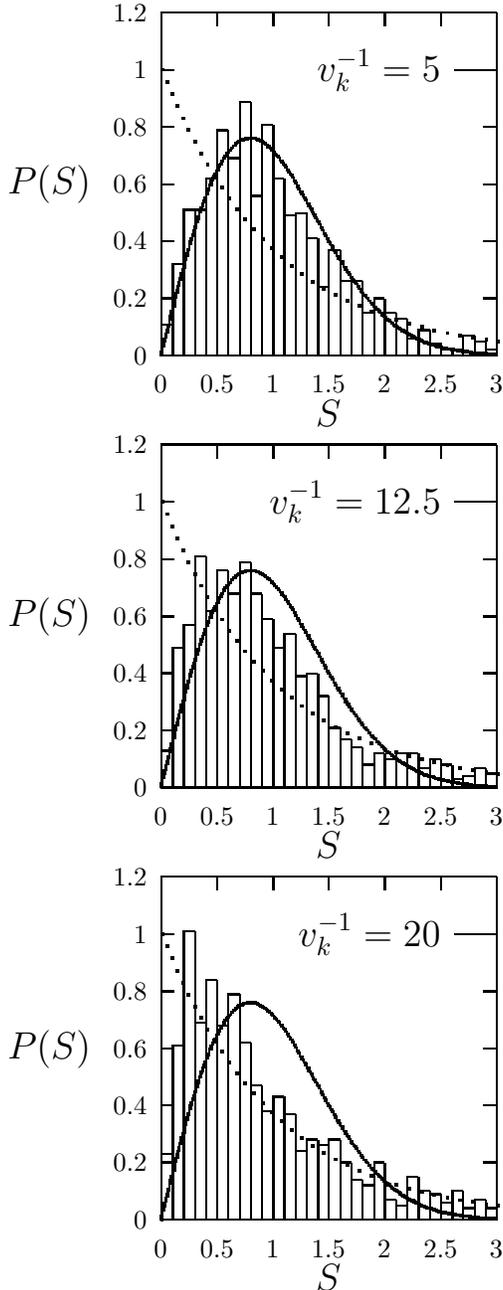

\input{fig1a.tex}
\input{fig1b.tex}
\input{fig1c.tex}
\caption{The dependence of the nearest-neighbor level spacing 
distribution on the strength of the point scatterers. 
The number of the scatterers is $N=5$ in all cases. 
The strength of all the scatterers is common. 
The statistics is taken among the eigenstates between 
$\omega_{100}$ and $\omega_{1100}$. 
The solid (broken) line is the prediction of GOE (Poisson statistics).}
\label{fig1}
\end{figure}
\begin{eqnarray}
\label{eq5-1}
E_{n_x,n_y} = \frac{1}{2M} \left\{
\left( \frac{n_x \pi}{l_x} \right)^2 + 
\left( \frac{n_y \pi}{l_y} \right)^2 \right\}, \\
\label{eq5-2}
\varphi_{n_x,n_y} (x,y) = \sqrt{\frac{4}{S}} 
\sin \frac{n_x \pi x}{l_x}
\sin \frac{n_y \pi y}{l_y}, 
\end{eqnarray}
respectively, ($n_x,n_y=1,2,\cdots$). 
We obtain $E_n$ and $\varphi_n$ by arranging 
the double-indexed eigenvalues and eigenfunctions in ascending 
order of magnitude of $E_{n_x,n_y}$. 
\begin{figure}[t]
\input{fig2.tex}
\caption{The dependence of the spectral rigidity on the strength of 
the point scatterers. The other indications are the same as in Fig.1.
}
\label{fig2}
\input{fig3.tex}
\caption{The dependence of the spectral rigidity on the number of 
the point scatterers. The strength of all the scatterers is $v_k^{-1}=7.5$. 
}
\label{fig3}
\end{figure}
We put several point scatterers inside the rectangle. 
The location of point scatterers, $\vec{x}_k = (x_k,y_k)$, 
is shown in Table \ref{table1}.  
The first $N$ coordinates are used in case of the number 
of scatterers being $N$. 
It is expected from Eq.(\ref{eq4-11}) that the $k$-th scatterer has a 
substantial effect on energy spectrum if its strength satisfies 
\begin{eqnarray}
\label{eq5-3}
\left| v_k^{-1} - \ln \omega \right| \alt \frac{\pi^2}{2} = 4.9348.
\end{eqnarray}
This is in fact the case, as seen from 
Fig.\ref{fig1} 
where the nearest-neighbor level spacing distribution 
$P(S)$ for three values of the strength of the scatterers is shown. 
The number of point scatterers is $N=5$ in all cases. 
The strength of all the scatterers is common.  
The statistics is taken within one thousand eigenvalues between 
$\omega_{100}=110.3579$ and $\omega_{1100}=1138.9682$. 
In this energy region, the chaotic spectrum is expected to appear for 
$v_k^{-1} \simeq 5 \sim 7$. 
The coincidence with the GOE prediction (solid line) is 
satisfactory for $v_k^{-1}=5$. As $v_k^{-1}$ increases, 
the level repulsion becomes weak and $P(S)$ shows intermediate 
between Poisson and GOE for $v_k^{-1}=12.5$,  
as expected from Eq.(\ref{eq5-3}). 
A further increase of $v_k^{-1}$ makes the distribution 
approach the Poisson distribution (broken line), 
though there still remains the level repulsion even for $v_k^{-1}=20$.  
A similar tendency can be observed in 
the spectral rigidity $\Delta_3 (L)$, which 
Fig.\ref{fig2} 
shows for several values of $v_k^{-1}$ under the 
same condition as in 
Fig.\ref{fig1}. 
The $\Delta_3 (L)$ statistics is close to 
the GOE prediction (solid line) in case of 
$v_k^{-1} = 5$ or $7.5$. 
As $v_k^{-1}$ increases, the gradual approach to 
the Poisson statistics (broken line) is observed. 
There still remains a considerable difference 
from the GOE prediction even in case of the maximal 
coupling. However, it tends to disappear as the number 
of scatterers increases. 
Figure \ref{fig3} 
shows the dependence of the spectral rigidity 
on the number of the scatterers. 
The strength of all the scatterers is $v_k^{-1}=7.5$, 
which is close to the maximal coupling strength. 
One can observe the gradual shift to the GOE prediction 
as the number of the scatterers increases. 
 
\section{Conclusion} 

We have discussed the condition for the appearance of wave chaos 
in the two-dimensional quantum pseudointegrable billiards 
with multiple point scatterers inside. 
Chaotic spectrum appears if the condition of Eq.(\ref{eq4-11}) is satisfied. 
It is described by a logarithmically energy-dependent strip with 
a constant width. 
The validity of our prediction is confirmed by numerical experiments  
with a rectangular billiard which contains multiple point obstacles 
inside. 
The degree of chaos depends on the number of the scatterers. 
We observe the GOE-like spectrum for $N=10$ scatterers with 
the maximal coupling strength. 

It should be stressed that, 
besides a fundamental aspect as a dynamical system, 
the quantum billiard with point interactions has a close 
relation to real systems. 
The quantum billiard is a natural starting point for 
examining the particle motion in microscopic bounded regions. 
The rapid progress in the microscopic or mesoscopic 
technology makes it possible to realize such settings. 
Real systems are, however, not free from small impurities 
which affect the particle motion inside. 
In the presence of a small amount of contamination, 
even a single-electron problem becomes unmanageable 
in an analytic manner. 
The modeling of the impurities with point interactions 
makes the problem easy to handle without changing 
essential dynamics. 
We hope that the current work serves as a guideline 
of the future research on the effect of impurities 
in mesoscopic devices.

\newpage
\vspace{30mm}
{\bf Profile of the Authors}

\vspace{10mm}

Takaomi Shigehara

email: sigehara@ics.saitama-u.ac.jp

http://www.me.ics.saitama-u.ac.jp/\~{ }sigehara/\\
{\it received the B.S, M.S and Ph.D. degrees in Physics 
from the University of Tokyo in 1983, 1985 and 1988, 
respectively. 
He is currently an Assistant Professor in the Department of 
Information and Computer Sciences at Saitama University. 
His research interests are quantum chaos, 
high-performance computing, and numerical analysis. 
}

\vspace{10mm}

Hiroshi Mizoguchi

email: hm@ics.saitama-u.ac.jp

http://www.me.ics.saitama-u.ac.jp/\~{ }hm/\\
{\it received the B.E. degree in Mathematical Engineering in 1980, 
and the M.E. and Ph.D. degrees in Information Engineering 
in 1982 and 1985, respectively, from the University of Tokyo.  
He is currently an Associate Professor in the Department of 
Information and Computer Sciences at Saitama University. 
His research interests are vision processor system, robotics, and  
quantum chaos. 
} 

\vspace{10mm}

Taketoshi Mishima

email: misima@ics.saitama-u.ac.jp

http://www.me.ics.saitama-u.ac.jp/\~{ }mishima/\\
{\it received the B.E., M.E. and Ph.D. degrees in 
Electrical Engineering 
from Meiji University, Japan, in 1968, 1970 and 1973, 
respectively. 
He is currently a Professor in the Department of 
Information and Computer Sciences at Saitama University. 
His research interests are foundation of 
symbolic and algebraic computation, 
axiomatic logic system, mathematical pattern recognition, 
and quantum chaos.}

\vspace{10mm}

Taksu Cheon

email: cheon@mech.kochi-tech.ac.jp

http://www.kochi-tech.ac.jp/\~{ }cheon/\\
{\it received the B.S., M.S. and Ph.D. degrees in Physics 
from the University of Tokyo in 1980, 1982 and 1985, 
respectively. 
He is currently an Associate Professor in the 
Laboratory of Physics at Kochi University of Technology.  
His research interests are quantum mechanics, chaos, and 
quantum chaos.}

\end{document}